\documentclass[%
 reprint,
 amsmath,amssymb,
 aps,
 showkeys,
]{revtex4-2}

\usepackage[dvipdfmx]{graphicx}
\usepackage{dcolumn}
\usepackage{bm}

\usepackage{color}
\definecolor{Red}  {rgb}{1,0,0}
\definecolor{Green}{rgb}{0,1,0}
\definecolor{Blue} {rgb}{0,0,1}
\usepackage{bm}
\newcommand {\bfv}[1] {{\boldsymbol {#1}}}
\newcommand {\ds} {\displaystyle}
\newcommand {\IND} {\hspace*{10pt}}
\newcommand\Rey{\mbox{{\rm Re}}}     
\newcommand {\com}[1] {}
\newcommand {\japanese}[1] {}
\usepackage{cancel}

\begin{document}

\preprint{APS/123-QED}

\title{Probabilistic description of flake orientation suspended in rotating wave flows}


\author{Tomoaki Itano}
\email{itano@kansai-u.ac.jp}
\affiliation{
  Department of Pure and Applied Physics,
  Faculty of Engineering Science, Kansai University,  Osaka, 564-8680, Japan
}
\author{Isshin Arai}
\affiliation{
  Department of Pure and Applied Physics,
  Faculty of Engineering Science, Kansai University,  Osaka, 564-8680, Japan
}

\date{\today}

\begin{abstract}
  In fluid dynamics experiments, flake-based flow visualization is a common technique to capture flow structures through the rays reflected from flat tracers suspended in the fluid.
  However, the correspondence between light intensity patterns in visualization images and the underlying physical properties of the flow can only be elucidated when the flow is known {\it a priori}.
  To reframe this limitation, just as the introduction of spin variable transformed quantum mechanics, we introduced the orientation variable into fluid dynamics and derived the time-dependent equation of the tracer orientation probability density field from an Eulerian perspective. 
As a first example in which a dimensionless parameter distinguishes the dependency on the initial condition, we illustrated an analytical solution of the orientation probability in a rotating wave flow.
With the inclusion of the diffusion term in the governing equation, the probability converged to the flow-determined state with spatially varying anisotropy, eliminating dependency on initial conditions.
As a second example, we solved the orientation probability field in the axisymmetric state in spherical Couette flow, to demonstrate independence from initial conditions consistent with experimental observations.
An asymmetric pattern in experimental images, unexplained by the dynamics of the tracer orientation, was reproduced from the unique solution of the proposed equation.
\end{abstract}


\maketitle


\section{Introduction}
Flow visualization using flat reflective tracers has long been used to capture flow structures in experiments, where fine reflective tracers suspended in the fluid, such as mica or aluminum flakes, align with the flow shear while maintaining a uniform number density\cite{Van82}.
Jeffery showed that the orientation of an axisymmetric particle suspended in a viscous shear flow follows a periodic trajectory on the unit sphere\cite{Jef22}.
For thin flat tracers, equilibrium orientations emerge on the Jeffery orbit. This alignment causes reflection of incident light to a specific direction, revealing spatial variations in the flow's shear field.
However, the correspondence between the light intensity patterns in visualization images and the underlying physical properties of the flow can only be elucidated when the flow is known {\it a priori}, as follows.

Gauthier et al.\cite{Gau98} calculated the flow patterns visualized by reflective tri-axial ellipsoidal tracers in three-dimensional flows and compared them with experimental results from Taylor–Couette flow and the shear flow between rotating disks. Based on the close agreement between their simulations and experiments, they concluded that Brownian motion and tracer history play only secondary roles in these steady flows.
They further noted that, in more general flow fields, there is no straightforward answer to what is actually being visualized, making it impossible to reconstruct the velocity field solely from the observed patterns.

By reflective flat tracers, Goto et al.\cite{Got11} experimentally captured sharp circles within a precessing spherical container.
Their numerical simulations of rotational dynamics of tracers in the flow showed that the experimentally observed circles correspond to regions with an isotropic orientation distribution.
When time is scaled by the Kolmogorov timescale, orientation dynamics collapse across the Reynolds numbers, revealing strong influence from small-scale turbulence.
Kida\cite{Kid14} pointed out that the visualized patterns are not solely due to the local flow, but rather arise from the overall contribution of the velocity gradient over the whole orbit of flat tracers consisting of the conical shear layers in a precessing sphere, and suggested that Brownian motion playing a key role in the orientation.

Two thought experiments on flat tracer orientation in steady flows lead to contradictory expectations.
In flows that are steady under appropriate coordinate transformations, once tracers have aligned, it seems unlikely they would return to isotropy, raising the question of whether thermal effects are requisite for re-isotropization.
Yet, in an opposing view, isotropy could theoretically be restored in time-reversed flow without thermal effects.
This speculative incongruity behind isotropy restoration has motivated us to focus on tracer orientation dynamics specifically in quasi-steady flows, where this fundamental paradox naturally emerges.
In the present study, we establish a probabilistic description\cite{Sav85,Kha00,Kid14} of orientation density field on flat tracers advecting in flows, where the diffusion effect reconciles the theory with experimental images that show little sensitivity to initial conditions.

The remainder of this paper is organized as follows:
We first formulate the time-dependent equation of the orientation probability field for flat tracers suspended in flows from Eulerian perspective.
This framework bridges visualization images and the time-developing orientation probability field.
Next, we analyze a rotating wave flow showing apparent decorrelation from initial orientation without diffusion.
Our probabilistic approach reproduces asymmetric patterns in spherical Couette flow unexplainable by tracer dynamics alone.
Finally, we conclude with brief remarks summarizing our findings.

\section{Formulation}
Hereafter, the term 'flakes' denotes flat and neutrally buoyant tracers that are accurately advected by the instantaneous flow field of an incompressible fluid. Being infinitesimally small, the flakes do not perturb the flow, thereby serving as ideal passive markers for flow visualization.
An infinitesimal material lines $\delta\bfv{l}(t)$ passing through the position $\bfv{r}=\bfv{\xi}(t)$ in the flow at time $t$ evolves following $\delta\dot{\bfv{l}}=\delta\bfv{l}\cdot\bfv{\nabla}_{\bfv{r}}\bfv{u} + {\rm O}(|\delta\bfv{l}|^2)$\cite{Bat67}, where the velocity field $\bfv{u}$ at the position $\bfv{\xi}(t)$ also satisfies $\bfv{u}(\bfv{\xi}(t),t)=\dot{\bfv{\xi}}$ and $\bfv{\nabla}_{\bfv r}=\sum_{i=x,y,z}\bfv{e}_i {\partial_i}$.
The time variation of the infinitesimal surface element formed by two independent lines on a flake, $\bfv{S}(t)=\delta\bfv{l}_1\times\delta\bfv{l}_2$, obeys the equation $\dot{\bfv{S}}=-\bfv{\nabla}_{\bfv{r}}\bfv{u}\cdot\bfv{S}$, from which the governing equation of the unit normal vector of the flake, $\bfv{s}=\bfv{S}/|\bfv{S}|$, was previously derived in Ref.\cite{Got11},
\begin{equation}
  \dot{\bfv{s}}=\bfv{s}\times\bfv{s}\times\bfv{\nabla}_{\bfv{r}}(\bfv{u}\cdot\bfv{s}) \ \ .
  \label{eq:dynamics}
\end{equation}

\IND
As the spin variable, two-valued label, was introduced into quantum mechanics for the wave function when the spin of an electron became observable, we now introduce a new coordinate, continuous orientation variable $\bfv{s}$, into fluid dynamics -- a discrete-vs-continuous distinction highlighting the fundamental difference between quantum and classical field descriptors, despite their conceptual similarity.
The {\it orientation probability density field} from Lagrangian viewpoint, $P(\bfv{s},t;\bfv{r}_0)$, is the probability density that the flake located at a position $\bfv{r}_0$ at the initial time $t=0$ is oriented at time $t$ per a unit steradian locally around $\bfv{s}$.
This field virtually underlies the system, with flakes serving as physical tracers that make this field observable.
The local velocity gradient $\bfv{\nabla}_{\bfv{r}}\bfv{u}$ near the location $\bfv{\xi}(t)$ passed by a flake dictates the time evolution of its orientation by Eq.~(\ref{eq:dynamics}).
Thus, provided that the velocity field $\bfv{u}(\bfv{r},t)$ is given, the orientation dynamics can be deterministically obtained in a Lagrangian framework based on the above equation.

\IND
On the other hand, we may denote $P(\bfv{s};\bfv{r},t) ds^2$ from Eulerian viewpoint as the orientation probability field that the flake located at a position $\bfv{r}$ at time $t$ is oriented within an infinitesimal steradian $ds^2$ around $\bfv{s}$ on the unit sphere.
We emphasize that the conventional flake-based flow visualization technique in fluid dynamics experiments represents $P(\bfv{s};\bfv{r},t)$, which is an Eulerian observation at a fixed spatial location, in contrast to the Lagrangian perspective, the observation of $P(\bfv{s},t;\bfv{r}_0)$ which follows the trajectory of a fluid particle through space and time.
Taking into account the equivalence between the Lagrangian and Eulerian orientation probability densities, $P(\bfv{s},t;\bfv{r}_0)=P(\bfv{s};\bfv{\xi}(t;\bfv{r}_0),t)$, we will formulate the time-developing equation of $P(\bfv{s};\bfv{r},t)$ by deriving Eulerian description from the counterpart introduced along the trajectories of Lagrangian-tracked flakes.

\IND
Note the velocity $\dot{\bfv{\xi}}$ of a flake passing at $\bfv{\xi}(t;\bfv{r}_0)$ at time $t$ equals Eulerian velocity field $\bfv{u}(\bfv{r},t)$ at the same position, and the position of the flake is $\bfv{r}+\bfv{u}(\bfv{r},t)\Delta t$ at time $t+\Delta t$.
From an Eulerian viewpoint, the increment between the orientation probabilities at an orientation $\bfv{s}$ for the moving flake from $t$ to $t+\Delta t$ 
 is represented $P(\bfv{s};\bfv{r}+\bfv{u}(\bfv{r},t)\Delta t,t+\Delta t)-P(\bfv{s};\bfv{r},t)=\bigl[{\partial_t}+\bfv{u}\cdot{\bfv{\nabla}_{\bfv r}}\bigr]P(\bfv{s};\bfv{r},t) \Delta t$.
The right hand side equals to ${\partial_t}P(\bfv{s},t;\bfv{r}_0)\Delta t$, which is the Lagrangian increment of orientation probability density from $t$ to $t+\Delta t$ along the trajectories of the same Lagrangian-tracked flake.

\IND
From a Lagrangian perspective, the probability $P(\bfv{s},t;\bfv{r}_0)$ is conserved on the unit sphere, $|\bfv{s}|=1$, when measured along tracer trajectories, 
 so that the change in $P(\bfv{s},t;\bfv{r}_0)$ over a time interval $\Delta t$ can be calculated from the net flux across the boundary of a steradian $ds^2$ around $\bfv{s}$ on the unit sphere.
The probability flux is the product of velocity and probability, $\dot{\bfv{s}}P(\bfv{s},t;\bfv{r}_0)$, so that the net flux satisfies the probability conservation on the sphere, 
  $\int ds^2 \bigl\{ P(\bfv{s},t+\Delta t;\bfv{r}_0)-P(\bfv{s},t;\bfv{r}_0) \bigr\} = -\Delta t \oint d\bfv{s} \cdot \bigl(\dot{\bfv{s}}P(\bfv{s},t;\bfv{r}_0)\bigr)$,
where $d\bfv{s}$ in the integral on the right-hand side denotes the product of an infinitesimal line element and the outward-pointing normal vector on the unit sphere, that are taken along the boundary loop.

\IND
A comparison of first-order terms in $dt$, together with the application of Gauss's divergence theorem, leads to the derivation of the following equation: ${\partial_t}P(\bfv{s},t;\bfv{r}_0)+\bfv{\nabla}_{\bfv{s}}\cdot \bigl(\dot{\bfv{s}}P(\bfv{s},t;\bfv{r}_0)\bigr) = 0 $.
Differentiation $\bfv{\nabla}_{\bfv{s}}$ is performed with respect to the orientation variable $\bfv{s}$,
 whose endpoint is restricted to the unit sphere.
 We finally retrieve the equivalence between the Lagrangian and Eulerian orientation probability densities, then derive the Fokker-Planck equation for Eulerian orientation probability density\cite{Doi86}; $\ds {\partial_t}P+\bfv{\nabla}_{\bfv{r}}\cdot\bigl(\bfv{u}P\bigr)+\bfv{\nabla}_{\bfv{s}}\cdot\bigl(\dot{\bfv{s}}P\bigr)=0$.
Provided an expression of $\dot{\bfv{s}}$ determined by $\bfv{s}$ and $\bfv{u}$, we can calculate the orientation probability density $P(\bfv{s};\bfv{r},t)$ of the flake located at $\bfv{r}$ and oriented to $\bfv{s}$ at time $t$. 
For instance, if the infinitesimally thin, rod-like elements are suspended in the flow, their time evolution can be described as $\dot{\bfv{s}}=-\bfv{s}\times\bigl(\bfv{s}\times(\bfv{s}\cdot\bfv{\nabla}_{\bfv{r}}\bfv{u})\bigr)$.
The tangent vector field $\dot{\bfv{s}}$ defined on the unit sphere does not satisfy the incompressibility condition, $\bfv{\nabla}_\bfv{s}\cdot\dot{\bfv{s}}\ne 0$.

\IND
On the unit sphere of a Lagrangian-tracked tracer at time $t$ and location $\bfv{r}$, thermal fluctuations in orientation space facilitate the exchange of the probability between neighboring orientations $\bfv{s}$ and $\bfv{s}+\Delta\bfv{s}$.
This diffusion process drives a net flux from higher to lower probability densities, characterized by a diffusion coefficient $D_{\bfv{s}}$.
From Eulerian perspective, the diffusion process of probability between neighboring locations $\bfv{r}$ and $\bfv{r}+\Delta\bfv{r}$ in physical space is characterized by a diffusion coefficient $D_{\bfv{r}}$.
With the aid of the incompressible flow condition and Eq.~(\ref{eq:dynamics}), these diffusion-induced terms are incorporated as corrections to the previously discussed fluxes $\bfv{u}P$ and $\dot{\bfv{s}}P$, leading to the following modified form of the equation.
\begin{equation}
{\partial_t}P+\bfv{u}\cdot \bfv{\nabla}_{\bfv{r}}P+\bfv{\nabla}_{\bfv{s}}\cdot\bigl(P \bfv{s}\times\bfv{s}\times\bfv{\nabla}_{\bfv{r}}\bfv{u}\cdot\bfv{s}\bigr) = D_{\bfv{r}} \bfv{\nabla}_{\bfv{r}}^2 P + D_{\bfv{s}}\bfv{\nabla}_{\bfv{s}}^2 P  
\label{eq:P}
\end{equation}
It is moreover required that the integral $\int P(\bfv{s};\bfv{r},t) ds^2=1$ for any $\bfv{r}$ and $t$, and appropriate boundary conditions provided.
Our model predicts reflection patterns in both steady and turbulent flows.
This equation applies to various flow regimes and particle shapes when $\dot{\bfv{s}}$ is given, as shown for rod-like elements.
Interpreting $\int P ds^2$ as the variable concentration field have allowed one to consider the velocity field depending on the observable $\bfv{s}$ via $P$ as its feedback effects incorporating the fluid motion provide rich phenomena in active biological fluid systems\cite{Sai08}.

\section{Dependency on initial condition}
In general, the flow exhibits shear that varies spatially and temporally, preventing the orientation vector from converging.
In contrast, in a simple Couette flow, the flake orientation $\bfv{s}$ asymptotically aligns with the principal eigenvector of the velocity gradient tensor.
This corresponds to the orientation probability, even starting from an arbitrary initial distribution such as isotropic, collapsing into a delta function over time, reflecting perfect alignment. 
One may then ask whether, specifically in the case of in quasi-steady flows, the flake orientation necessarily converges to perfect alignment.
A rotating wave flow, analyzed through characteristic time scales, challenges this intuition.

\IND
Imagine a flake traversing an irrotational flow field.
If we assume that the velocity gradient perceived by the flake is small from a Lagrangian perspective, then the flake orientation exponentially approaches the direction of the eigenvector associated with the smallest strain eigenvalue due to the linearlity in Eq.~(\ref{eq:dynamics}).
However, in general such assumptions break down, so that the direction of the preferred eigenvector varying spatially is expected to change as the flake traverses the field.
The time constant required for the flake to align is estimated as the inverse of the absolute value of the smallest (negative) real component among the eigenvalues of the velocity gradient tensor, that is, $ 1/t_{\rm align} \sim |{\rm min}({\rm E.V. of } \bfv{\nabla u})|$.
In areas of spatially varying velocity gradients, the orientation of the preferred eigenvector changes significantly within a short distance for which flakes traverse during the interval required for the orientation probability to localize.
Suppose the preferred orientation at $\bfv{r}$ is represented by $\bfv{e}(\bfv{r})$ and the flake moves from $\bfv{r}$ to $\bfv{r}'=\bfv{r}+\int_0^{t_{\rm \tiny align}} \bfv{u}(\bfv{\xi}(t),t) dt$ in $t_{\rm \tiny align}$.   
In regions where the condition $|\bfv{e}(\bfv{r})\cdot \bfv{e}(\bfv{r}')|<1$ is realized, the flake orientation remain insufficiently aligned along the short path.
A dimensionless parameter as the ratio of two relevant time scales is envisaged, which serves to indicate whether perfect alignment is expected.

\IND
As the first illustrative case with respect to such a dimensionless parameter, we consider a two-dimensionally quasi-steady rotating wave flow in which the orientation of the shear rotates about the origin with constant angular velocity $\omega_0$.
The flow is generated from the streamfunction $\Psi(\bfv{r})=-(\bfv{n}(t)\cdot\bfv{r})^2 {\Omega_0}/{2}$, where $\bfv{n}(t)=\bfv{e}_x\cos{\omega_0 t}+\bfv{e}_y\sin{\omega_0 t}$, $\Omega_0$ and $\omega_0$ are positive\cite{Yos23}.
Note that the streamlines differ from the flake trajectories in the stationary frame of reference because of $\partial_t\bfv{u}\ne 0$ and that this flow corresponds to the simple Couette flow at $\omega_0\to 0$.
The time-varying shear experienced by advecting flakes in general quasi-steady flows can be effectively represented by a stationary flake at the origin in this specific rotating wave flow.
When diffusion is absent, Eq.~(\ref{eq:P}) possesses the analytical time-dependent solution with respect to a flake staying at the origin, which was solved by examining the method of characteristics.
Orientation probability in the rotating frame of reference, $\beta$ (the angle against $\bfv{n}$), collapses into a delta function for $\omega_0/\Omega_0 < 1$ with any initial condition $P(\beta;0)$, while it is permanently oscillative with a period $\pi/\sqrt{\omega_0(\omega_0-\Omega_0)}$ for $\omega_0/\Omega_0 > 1$ and dependent on the initial condition $P(\beta;0)$.
The solution has a number of conservative quantities, $\int\, (\Omega_0\sin^2{\beta'}-\omega_0)^n P^{n+1}(\beta';t) d\beta'$ for $n=0,1,2,\cdots$.
Although the system may seem to have lost correlation with its initial orientation, the presence of a conserved quantity in the absence of diffusion ensures that the initial condition is never completely erased; The apparent loss of correlation with the initial orientation is merely a seeming relaxation.
In the presence of diffusion, the system numerically converged to a steady state with spatially varying anisotropy in oscillatory decay even for $\omega_0/\Omega_0>1$, independent of initial conditions.
We next turn to spherical Couette flow as a second case where diffusion effect explains experimental visualization patterns.

\begin{figure}
  \centering
  \includegraphics[angle=0,width=0.90\columnwidth]{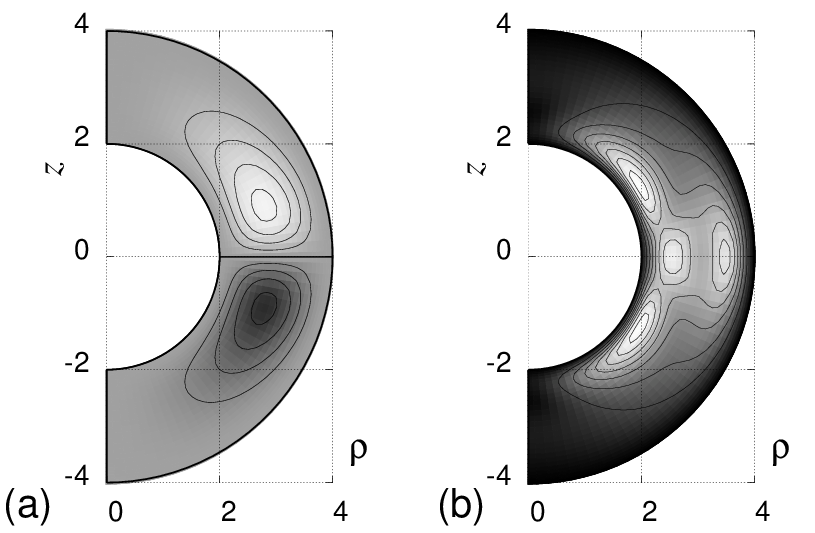}
  \caption{
    Axisymmetric state of the spherical Couette flow at $\Rey=80$ projected on a meridional plane.
    (a) Streamlines of the meridional circulation,
    (b) contour plot of the anisotropy defined as $\iint \bigl(P(\bfv{s};\bfv{r})-{1}/{4\pi}\bigr)^2 d|\bfv{s}|^2$ for diffusion coefficients $(D_\bfv{r},D_\bfv{s})=(1,1/16)$.
    Note the reduced anisotropy near the rotational center of the circulation.
  }
  \label{fig:stream}
\end{figure}

\begin{figure}
  \centering
  \includegraphics[angle=0,width=0.80\columnwidth]{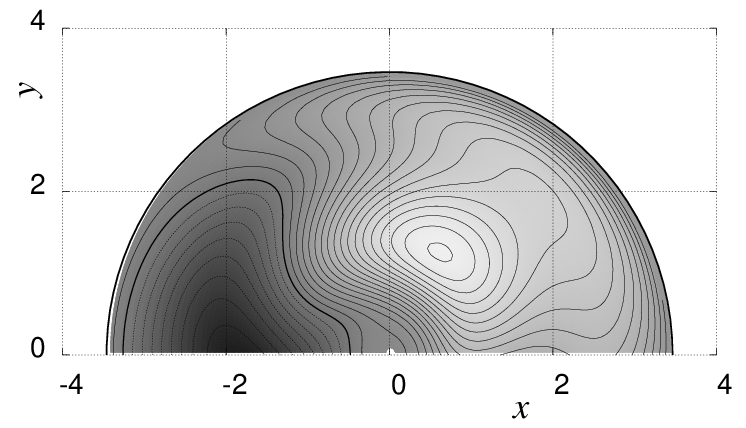}
  \caption
  {
    Simulated visualization of the axisymmetric state of spherical Couette flow as viewed from the polar side with the incident light parallel to the $y$-axis, calculated from the solution of Eq.~(\ref{eq:P}).
    The asymmetric pattern observed in experimental images, which contradicted the expectation based on the equilibrium solution of Eq.~(\ref{eq:dynamics}), where $\bfv{s}_{\pm}$ is perpendicular to $\bfv{e}_\phi$ and independent of $\phi$), was successfully reproduced by our probabilistic model including diffusion effects.
  }
  \label{fig:image}
\end{figure}

\section{Regression to isotropy}
The diffusion effect should reconcile the theory with the experimental evidence that practically shows little sensitivity to initial conditions.
Aluminum flakes have been utilized in experimental visualization studies of Spherical Couette flows\cite{Egb95,Yos23,Ara24}.
In axisymmetric steady flows, the resulting visualization images were highly reproducible; It is unlikely that the procedure to mix flakes into flows had an impact on the captured images.Let us examine this using Eq.~(\ref{eq:P}).

\IND
The steady axisymmetric flow consists of a primary equatorial jet $u_\phi(r,\theta)\bfv{e}_\phi$ encircling the rotation axis of the inner sphere and a meridional circulation $u_r(r,\theta)\bfv{e}_r+u_\theta(r,\theta)\bfv{e}_\theta$ looping between the equator and the poles (see Fig.\ref{fig:stream}(a)); the latter is regarded as Ekman pumping induced near the boundary layers.
A streamfunction characterizes the meridional circulation, along which a flake follows a helical trajectory confined to a toroidal surface where the streamfunction remains constant.
From the deterministic viewpoint, Eq.~(\ref{eq:dynamics}) is satisfied by either $\bfv{s}\cdot\bfv{u}=0$ or $\bfv{s}\cdot\bfv{e}_\phi=0$, that is, each condition defines a time-invariant set of orientations for a flake moving along its helical trajectory.
The two invariant sets, which are great circles on the unit sphere, divide the unit sphere surface into the four regions that never mix permanently.
Particularly, the two points at which these invariant sets intersect correspond to exact solutions $\bfv{s}_{\pm}(t)$, which can be recognized as spatial fields in the axisymmetric state of spherical Couette flow.
Numerical integration of Eq.~(\ref{eq:dynamics}) with no Langevin force from arbitrary initial conditions $\bfv{s}(0)$ revealed $\bfv{s}(t)$ converges toward one of the two exact solutions $\bfv{s}_{\pm}(t)$, being structurally unstable degenerate saddle associated with the aforementioned apparent loss of correlation with the initial state\cite{Guk83}.

\IND
If flake orientations would converge to $\bfv{s}_{\pm}(t)$ in reality, which are perpendicular to $\bfv{e}_\phi$ and independent of $\phi$, the images captured from the polar side, under illumination by an incident light sheet perpendicular to the rotation axis, would be mirror-symmetric with respect to the plane passing through the rotation axis and parallel to the optical axis; in practice, however, experimentally obtained images deviated from the mirror symmetry.
To capture the complex interplay between spatial and orientation dynamics, we developed a novel approach using double spherical harmonics expansions.
Using the Crank-Nicolson method, we numerically solved Eq.~(\ref{eq:P}) 
 within a five-dimensional extended space comprising both physical and angular coordinates, under the Dirichlet boundary condition with finite diffusion coefficients.
The numerical solution converged to a unique solution guaranteed by the maximum principle applied to the second-order linear partial differential equation\cite{Gil77} involving a generalized elliptic operator.
Energy integral analysis showed that initial deviations from the unique steady solution decay asymptotically, with or without oscillatory behavior depending on parameter values.
For an inner-to-outer sphere radius ratio of $1/2$ and $\Rey=160$, we obtained the rays reflected to the polar direction when a light sheet was introduced from the $+y$ direction, perpendicular to the rotation axis, onto a plane passing through the inner sphere pole (see Fig.\ref{fig:image}).
The asymmetric pattern in experimental images, which was previously unexplained, was well reproduced using solutions from the proposed equations, by introducing Brownian motion to the present system.
Furthermore, in the vicinity of the rotational center of the meridional circulation, the intensity of shear is expected to be weak, making localization of orientation probability less likely as seen in Fig.\ref{fig:stream}(b), which was pointed out in Taylor-Couette flow\cite{Abc08}.

\section{Concluding remarks}
By introducing the orientation variable into fluid dynamics, we derived the time-dependent equation of the tracer orientation probability density field from an Eulerian perspective.
The probability field virtually underlies the system, with flakes serving as physical tracers that make the field observable.
The equation enables the prediction of reflected light images even in time-dependent turbulent flows,
 without solving the dynamics of each flake suspended in flows.
We presented an analytical example in rotational quasi-steady flows in which the initial orientation may persist within the probability distribution, and numerically demonstrated that diffusion not only promotes isotropization but also eliminates the dependence on initial orientation probability.
In the case of the axisymmetric solution of spherical Couette flow, our formulation was able to reproduce an asymmetric pattern that could not be predicted from dynamics alone.
The probabilistic description of flake orientation provides a fundamental link between observable patterns and the underlying flow physics,
 thereby offering a pathway to connect flake-based flow visualization images with essential physical properties of the flow.
For applications such as dense suspensions and biological flows, a natural extension would be to consider feedback mechanisms where the orientation field modifies the flow itself, bearing subtle parallels to concepts like spin-orbit interactions in quantum systems.

\begin{acknowledgments}
  The authors would like to thank
  Mr. K. Yoshikawa
  for his pilot experimental surveys.
  They also acknowledge ORDIST in Kansai University and the RIMS Joint Research Activities in Kyoto University for providing a space for their research and communication.
  This work was supported in part by a Grant-in-Aid for Scientific Research(C) and JSPS KAKENHI Grant No.24K07331.   
\end{acknowledgments}

\bibliography{scf12}

\end{document}